\documentstyle[twocolumn,epsfig]{IEEEtran}
\def\BibTeX{{\rm B\kern-.05em{\sc i\kern-.025em b}\kern-.08em
T\kern-.1667em\lower.7ex\hbox{E}\kern-.125emX}}

\begin{document}

\title{Modeling nonlinearity and hysteresis due to critical-state flux penetration 
in hard superconductors}

\author{T. Dasgupta, Durga P. Choudhury and S. Sridhar \\
{\small Department of Physics, Northeastern University, Boston, MA 02115}}
\maketitle

\begin{abstract}
The nonlinear surface impedance of a thin superconducting strip carrying a
microwave current has been calculated numerically from first principles
based upon flux penetration due to a current-induced critical state. The
calculations approach analytic results in the appropriate limit. The
technique is useful for design of passive superconducting microwave circuits.
\end{abstract}


\section{Introduction}

\PARstart{S}{ince} its discovery, the new breed of oxide superconductors
with transition temperatures above the boiling point of nitrogen have shown
great potential as a replacement for metals for passive microwave structures
owing to their very low surface resistance\cite
{DPChoudhury98a,MJLancaster,ZYShen94b}. However, a key feature of the HTSC
is the inherently nonlinear response at microwave frequencies, which is
present over a wide range of input powers. This nonlinear responses also
poses a major challenge for numerical device design using these materials.
Most existing computer aided design software do not take into account the
inherent nonlinearity of superconductors. The few analytical models that do
exist\cite{SSridhar94a} are valid only for highly idealized geometries and
need to be incorporated into practical device design.

There are several physical models of nonlinear surface resistance, each
valid in a different regime of temperature and microwave power\cite
{MGolosovsky95a}. The most pursued ones are pinning induced critical state%
\cite{SSridhar94a,JMcDonald97a,JMcDonald98a}, transport current induced
self-heating \cite{Golosovsky}, and depairing \cite{Golosovsky}. The current-driven
critical state model \cite{SSridhar94a} has proven to be particularly useful
not only because it fits experimental measurement of $R_{S}$ very accurately
over a wide range of temperature and power levels \cite{BAWillemsen95a}, but
also because its usefulness extends to modeling hysteretic losses in
superconducting transmission lines.

The (field-induced) critical state model was first proposed by Bean\cite
{CPBean62a} to explain the unusual magnetization behavior of hard
superconductors and was first solved for the geometry of a semi-infinite
slab where the field was applied parallel to its face. The geometry where
the field is applied perpendicular to a thin strip was solved relatively
recently by several groups and the field and current distribution\cite
{EZeldov94c,EHBrandt93a} and the surface impedance from hysteretic loss\cite
{SSridhar94a} was obtained analytically.

Though very elegant, the analytic results were restricted to certain
``ideal'' geometries only where symmetry could be used to simplify the
problems. However, these idealizations are not always satisfied in real
world applications of superconducting strips in microwave circuits. In the
wake of the fact that the model does describe experimental data accurately 
\cite{BAWillemsen95a}, it is imperative to be able to implement a numerical
paradigm without the constraints of the idealized geometry required by the
analytic approach.

In this paper, we demonstrate a numerical approach to incorporate the
nonlinear response of superconducting materials using assumptions equivalent
to those of the current induced critical state.

\section{The Critical State Model}

\begin{figure}[tbp]
\mbox{\epsfig{file=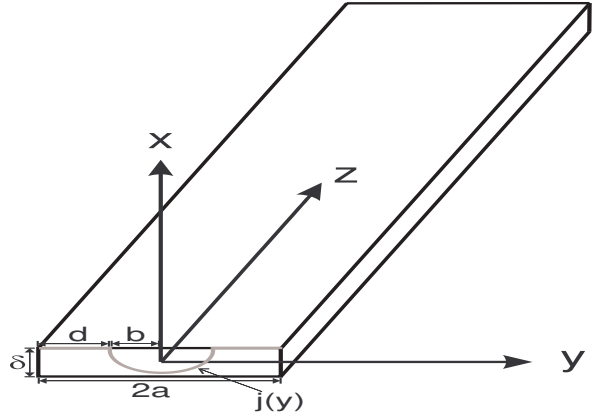, height=2.4 truein, width=0.45\textwidth}}
\caption{ The geometry of the problem. The strip is centered around $x=y=0$.}
\label{Fig1}
\end{figure}

Let us consider a stripline with  the ground plane  far away from the center
conductor so that it carries a  negligible current and hence does not effect the
magnetic field distribution inside the center conductor. It has been shown
in literature that such an assumption is reasonable\cite
{CampbellIEEE} (We will discuss the effects of ground planes somewhere
else.) The geometry of the center conductor and our choice of axes is shown
in of Fig.\thinspace \ref{Fig1}. 

Starting from the most generic physical assumptions that  retain the
essential ideas of critical states:

\begin{enumerate}
\item  $|{\bf j}({\bf r})|\leq j_c$ where ${\bf j}$ is the current density.

\item  The magnetic induction ${\bf B}(x,y)=0$ for $|y|\leq b\leq a$
\end{enumerate}
The first assumption highlights the fact that flux lines are pinned
until the Lorenz force on them exceeds a certain critical value. The
second assumption follows from the fact that the vortices start penetrating
from the edges. Hence for any given current there would be a flux free
region in the center of the strip. Note that the depinning critical current
is at least an order of magnitude smaller than the depairing critical
current where superconductivity would be quenched by the self field. 

In the following we show that for a strip with finite thickness, this model
can be solved numerically and that it leads to  unique  field and current
distribution within the strip. We further show that $(1)$ and $(2)$ leads to
hysteresis irrespective of the geometry  and study the hysteresis laws for
the specific strip geometry under consideration.

\section{Computational Details}

A semi-numerical approach was used to determine a current distribution which
satisfies the two assumptions presented above. Assuming an uniform current 
distribution along the x-direction one can show that
\begin{equation}
{\bf B}(x=0,y)=\frac{\mu _0}\pi \int_{-a}^a{\bf j}(y^{\prime })\tan
^{-1}[\frac \delta {2(y-y^{\prime })}]dy^{\prime }
\label{analyticalExp}
\end{equation}
The above analytical expression for $B(x=0,y)$ in Eq(\ref{analyticalExp})
was used to save computational time.

In the model which requires that ${\bf j}(y:b\leq |y|\leq a)\equiv J_c$
and $B(x=0,|y|\leq b<a)=0$, one needs to solve for the current distribution
in the field free region $|y|\leq b<a$ 
\begin{equation}
\int_{-a}^a{\bf j}(y^{\prime })\tan ^{-1}[\frac \delta {2(y-y^{\prime
})}]dy^{\prime }=0\;{\rm for\;}|y|\leq b<a  \label{IntegralEq}
\end{equation}

Numerical solutions of integral equations such as eq(\ref{IntegralEq}) are
rather complicated. We found a simpler iterative approach to find the
current density in the field free region of the superconducting strip. Here
we give a brief description of the algorithm that was used.

Suppose that the thickness of the strip is infinite ($\delta \rightarrow
\infty $.). Then for any given initial current distribution ${\bf j}(y)$, a
field free region $B(|$ $y|\leq b)\equiv 0$ can be obtained by replacing the
current density in that region by

\begin{equation}
j(|y|<b)\Rightarrow j(|y|)+\frac{\partial }{\partial y}B_{x}
\label{iteration1}
\end{equation}
Also note that the magnetic field strength on the line $x=0$ for a slab ($%
\delta \rightarrow \infty $.) is mostly effected by the current on and
around the line $x=0$. Therefore, for the case of a finite $\delta $, the
transformation in eq(\ref{iteration1}) will bring the field strength in the
region $|y|<b$ a little closer to zero. Hence an iterative implementation of
the above transformation of $J(y)$ in eq(\ref{iteration1}) will finally
converge bringing the magnetic field strength in the region $|y|<b$ to zero,
irrespective of the size of $\delta $.

In a nutshell, we start with a flat current distribution, ${\bf j}(y)=j_c$
in an one dimensional grid of size $4000$ and we iteratively make ${\bf B}%
(x=0,|y|<b)$ vanish by replacing the current density by $j_y^{n+1}%
\Rightarrow j_y^n+\left( B_{y+1}^n-B_y^n\right) /\Delta y$.

\section{Results}

\subsection{Penetration Law}

\begin{figure}[tbp]
\mbox{\epsfig{file=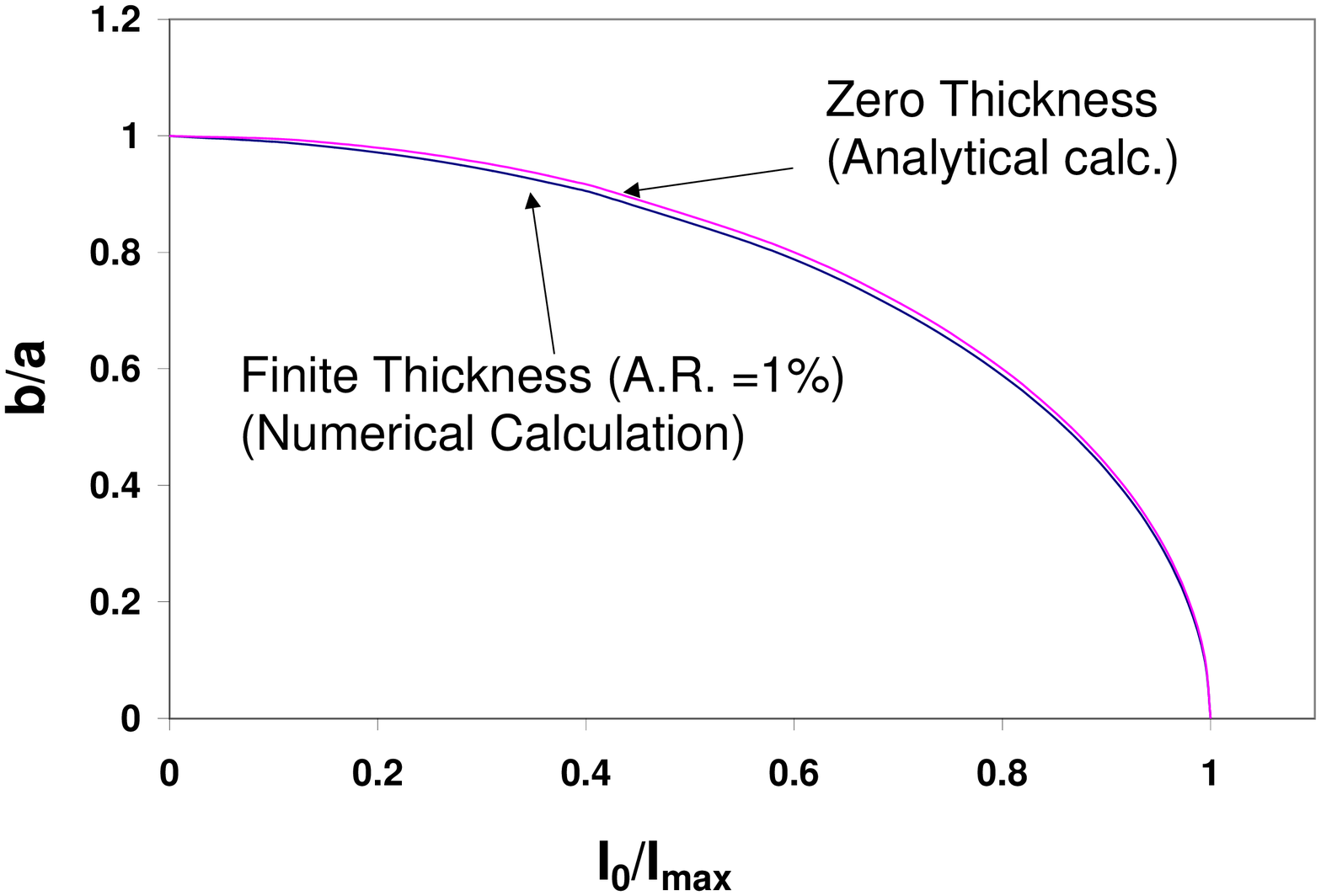, height=2.4 truein, width=0.45\textwidth}}
\caption{Penetration depth $b$ as a function of current $I/I_{\mbox{max}}$. }
\label{Fig2}
\end{figure}

\begin{figure}[tbp]
\mbox{\epsfig{file=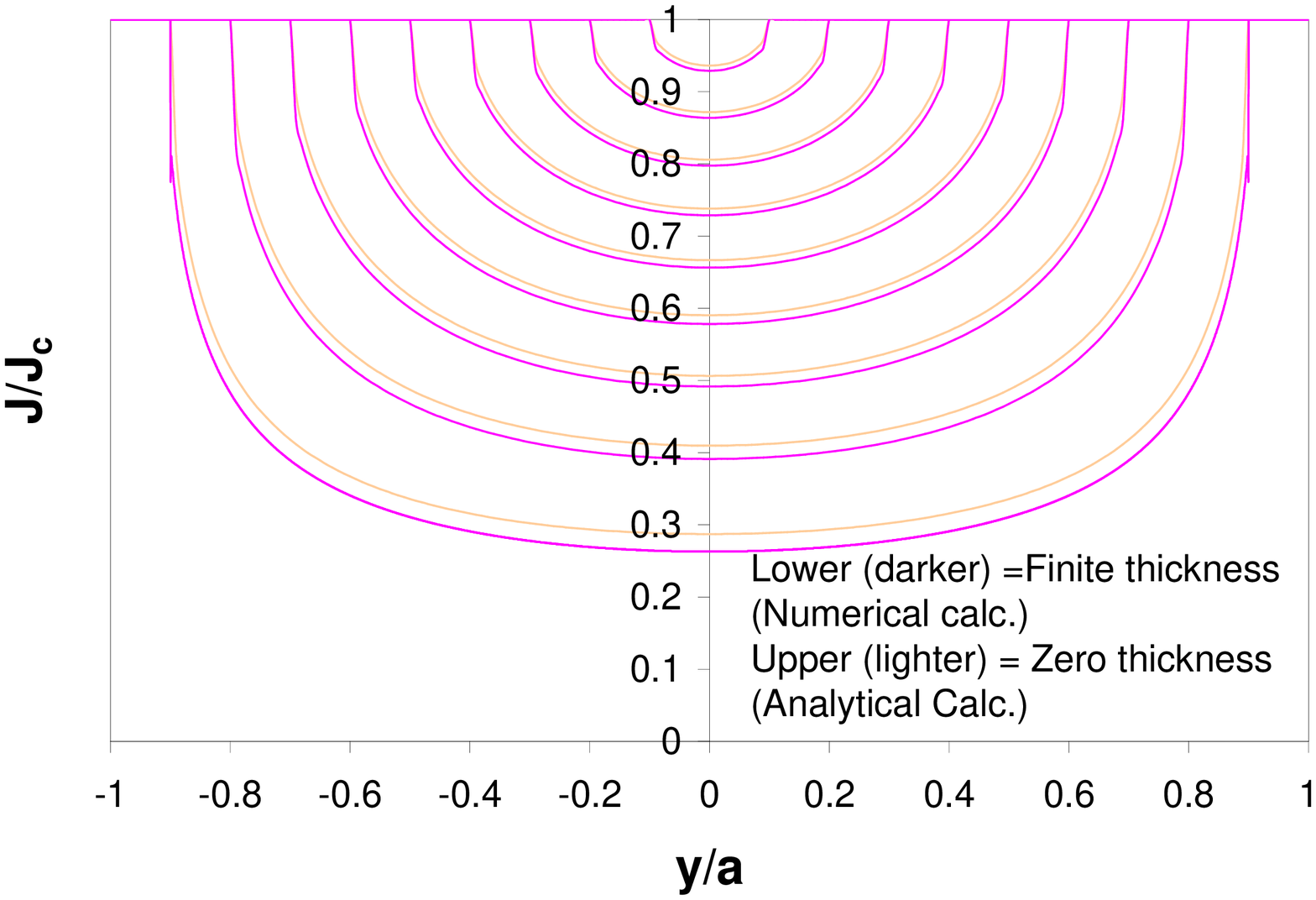, height=2.4 truein, width=0.45\textwidth}}
\caption{Current distribution $j/j_c$ across the strip for $b/a =
0.1,0.2,\ldots ,0.9$. The lower (solid) lines are numerical calculations and
the upper (dashed) lines are analytic calculations.}
\label{Fig3}
\end{figure}

\begin{figure}[tbp]
\mbox{\epsfig{file=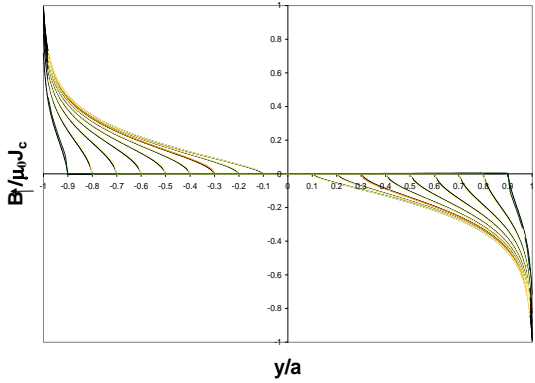, height=2.4 truein, width=0.45\textwidth}}
\caption{The magnetic field strength $B(x=0,y/a)$ across the strip for 
$I/I_{\rm max}= 0.41,0.58,0.70,0.79,0.86,0.91,0.95,0.98,0.99$.}
\label{Fig4}
\end{figure}

We calculate the flux penetration depth $d/a$ for a strip of aspect ratio of
1\% as a function of the applied current. In Fig.\ref{Fig2} we display $%
(1-d/a)=b/a$ as a function of the applied current $y/a$ and compare that
with the analytical results for a strip of zero thickness\cite{EHBrandt93a}.
Fig.\ref{Fig3} and Fig.\thinspace \ref{Fig4} shows the corresponding current
and field distributions. One can see from Fig.\ref{Fig2} and Fig.\ref{Fig3}
that the penetration law get's more linear and the current density in the
field free region decreases as we go from zero thickness to a finite
thickness. These results are expected since, in the limit $\delta
\rightarrow \infty $ we expect a linear penetration law and $J(x=0,|y|\leq
b<a)=0$. These results can then  determine the current critical states and
the corresponding magnetic field distribution for AC currents with a peak
current of $I_0$.

\section{Hysteresis}

The hysteresis law can be derived by studying the current density
distribution for an oscillating current over a full cycle. If the current
density distribution for a total current $I$ is $j(y,I,j_c\rightarrow\infty)$
then in absence of pinning the current density will be $Nj(y,I,j_c%
\rightarrow\infty)$ for a total current of $NI$. In other words 
\begin{equation}
j(y,NI,j_c\rightarrow \infty )=Nj(y,I,j_c\rightarrow\infty)
\end{equation}
However, since $j(y)\leq j_c$ everywhere, the current density distribution
depends on the total current. That is

\begin{equation}
j(y,NI,j_c)\neq Nj(y,I,j_c).
\end{equation}

When the current is reversed, after reaching the peak $I_0$, current density
starts to decrease from the edges with an effective critical current density
in the opposite direction of $2J_c$. This is because at the edges the
current density is $j_c$ in the positive x-direction.

\begin{equation}
j_{c\downarrow}=2j_{c\uparrow}
\end{equation}

\begin{figure}[tbp]
\mbox{\epsfig{file=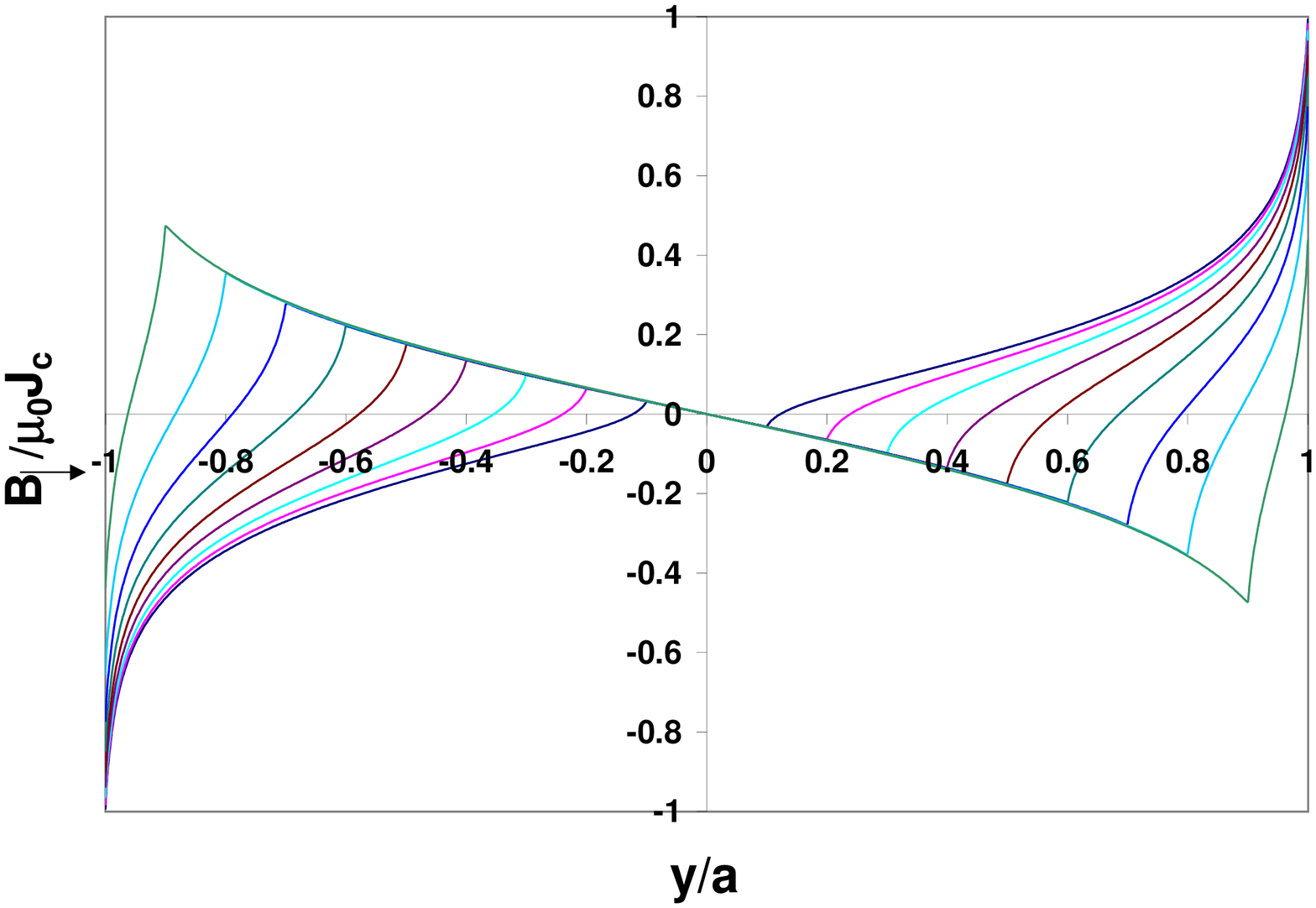, height=2.4 truein, width=0.45\textwidth}}
\caption{Magnetic field strength, $B(x=0,y/a)$, when $I$ is reduced from 
$I_{\mbox{max}}=2J_c\delta a$ to $-I_{\mbox{max}}$. This graph dispalys 
the origin of hysteresis. 
We plot $B$ across the strip for 
$I/I_{\rm max}=0.18,-0.17,-0.40,-0.58,-0.72,-0.82,-0.90,-0.96,-0.99$.}
\label{Fig5}
\end{figure}

Thus the hysteresis law on the downward cycle is given by\cite{EHBrandt93a} 
\begin{equation}
j_{\downarrow }(y,I,j_c)=j_{\mbox{max}}(y,I_0,j_c)-j_{\uparrow
}(y,I_0-I,2j_c)
\end{equation}
Fig.\thinspace \ref{Fig5} displays the magnetic field distribution when the
input current is reduced from the maximum current $2J_c\delta a$ to $%
-2J_c\delta a$. As  the decreases through $0$ one does not see any field
free region inside the strip, that is there are locked in fields from field
distribution when the total current was $2J_c\delta a$: something that is
unique to hysteresis.  

\section{Surface Impedance}

\begin{figure}[tbp]
\mbox{\epsfig{file=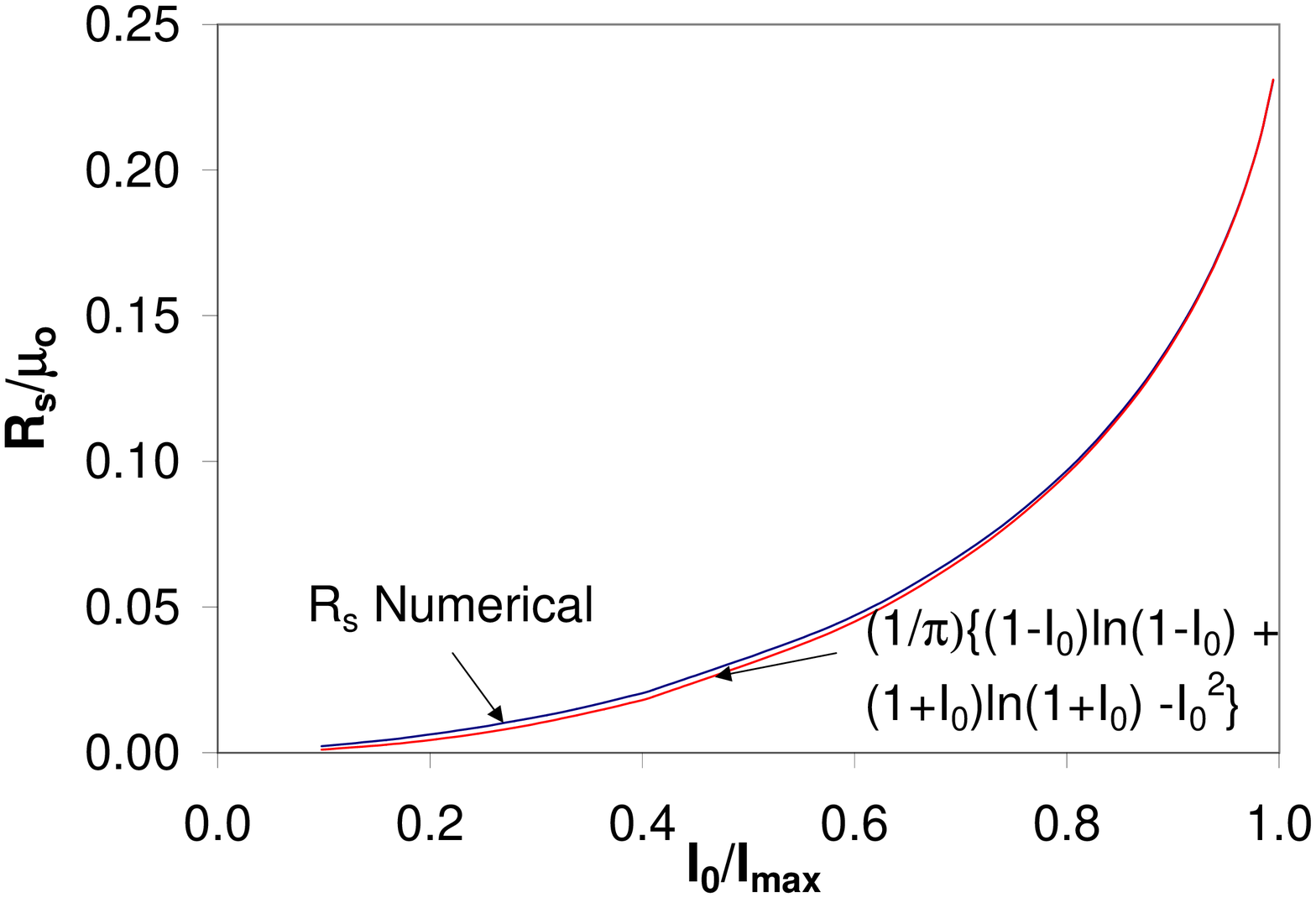, height=2.4 truein, width=0.45\textwidth}}
\caption{Surface resistance $R_s$ as a function of the current $I/I_{
\mbox{max}}$.}
\label{Fig6}
\end{figure}

Assuming that the strip is excited with a current of amplitude $I_0$ and
frequency $\omega_0$, the electric field $E(y,t)$ is be given by

\begin{equation}
E_{Z}(y,t)=\frac{d\ }{dt}\int_0^ady^\prime\ B_x(y^\prime,t)  \label{Efield}
\end{equation}
and the energy lost in a half cycle is 
\begin{equation}
U_{H.C.}=2\int_{0}^{a}dy\int_{-T/4}^{T/4}dt\;J(y,t)E(y,t)
\end{equation}
Substituting eq(\ref{Efield}) and integrating by parts we get 
\begin{eqnarray}
U_{H.C.} &=&2\int_{0}^{a}dy\int_{-T/4}^{T/4}dt\;J_{c}\frac{d\ }{dt}%
\int_{b(I_{0})}^{a}dy^{\prime }\ B_{x}(y^{\prime },t) \\
&=&4\mu _{0}J_{c}\int_{b(I_0)}^ady\int_{b(I_0)}^ady^\prime B_x(y^\prime ,t)
\end{eqnarray}
This definition of $U_{H.C.}$ was also used by Brandt in ref.\cite
{EHBrandt93a}. This defines the surface resistance as 
\begin{equation}
R_s=2\frac{4\mu_0J_c}{I_0^2T}\int_{b(I_0)}^ady\int_{b(I_0)}^ady^\prime
B_x(y^\prime ,t)
\end{equation}
Fig.\,\ref{Fig6} displays the surface resistance $R_s$ as a function of the
peak current, $I_0$.

The surface reactance $X_s$ per unit length can be calculated using the
average total electrostatic energy per unit length.

\begin{equation}
X_s=\frac 1{\mu _0I_0^2T}\int_0^T\int B^2(x,y)d^2s_{xy}
\end{equation}

\begin{figure}[tbp]
\mbox{\epsfig{file=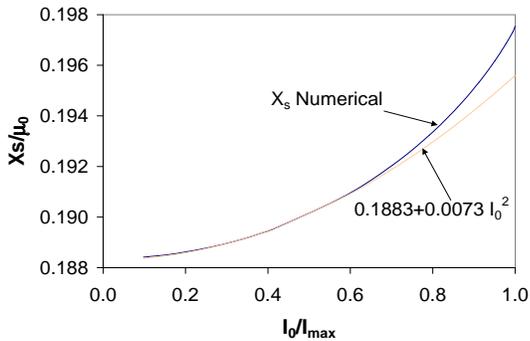, height=2.4 truein, width=0.45\textwidth}}
\caption{Surface reactance $X_s$ as a function of the current $I/I_{
\mbox{max}}$.}
\label{Fig7}
\end{figure}

We carry out the space integral until $B^2(x,y)$ falls down by $99\%$ of
it's peak value. Fig.\thinspace \ref{Fig7}, displays the surface reactance $%
X_s$ as a function of the peak current, $I_0$. Our results are close to the
analytical results in ref.\cite{SSridhar94a}.

\section{Conclusion}

A first principle numerical calculation on a strip of finite thickness using
assumptions similar to the critical state model yields results that approach
the analytic results in the appropriate limits $\left( \delta \rightarrow
0\right) $ but without the unrealistic idealizations of the later. Further a
numerical approach gives one the flexibility of accommodating higher
complexities like effects due to the ground planes and more accurate
calculations of surface impedance and harmonics. This topic will be addressed
in future work. This approach can be implemented into a
computer aided design framework for the design of passive superconducting
microwave components.

\section{Acknowledgments}

This work is supported by NSF-9711910 and AFOSR.


\begin{thebibliography}{11}

\bibitem{DPChoudhury98a}
Durga~P. Choudhury and S.~Sridhar,
\newblock ``Superconducting microwave technology,''
\newblock in {\em Wiley's Encyclopedia of Electrical and Electronic
  Engineering}, John Webster, Ed. John Wiley \& Sons, New York, 1999.

\bibitem{MJLancaster}
M.~J. Lancaster,
\newblock {\em Passive microwave device applications of high temperature
  superconductors},
\newblock Cambridge University Press, New York, 1997.

\bibitem{ZYShen94b}
Zhi-Yuan Shen,
\newblock {\em High-Temperature Superconducting Microwave Circuits},
\newblock Artech House, Boston, 1994.

\bibitem{SSridhar94a}
S.~Sridhar,
\newblock ``Non-linear microwave impedance of superconductors and {ac} response
  of the critical state,''
\newblock {\em Appl. Phys. Lett.}, vol. 65, no. 8, pp. 1054--1056, August 1994.

\bibitem{MGolosovsky95a}
M.~A. Golosovsky, H.~J. Snortland, and M.~R. Beasley,
\newblock ``Nonlinear microwave properties of superconducting {Nb} microstrip
  resonators,''
\newblock {\em Phys. Rev. B}, vol. 51, no. 10, pp. 6462--6469, March 1995.

\bibitem{JMcDonald97a}
J.~McDonald, John~R. Clem, and D.~E. Oates,
\newblock ``Critical-state model for harmonic generation in a superconducting
  microwave resonator,''
\newblock {\em Phys. Rev. B}, vol. 55, no. 17, pp. 11823--11831, May 1997.

\bibitem{JMcDonald98a}
J.~McDonald, J.~R. Clem, and D.~E. Oates,
\newblock ``Critical-state model for intermodulation distortion in a
  superconducting microwave resonator,''
\newblock {\em J. Appl. Phys.}, vol. 83, no. 10, pp. 5307--5312, May 1998.

\bibitem{Golosovsky}
M.~ Golosovsky and M.~Tsindlekht and H.~Chayet and D.~Davidov
\newblock ``Vortex depinning frequency in {$\rm YBa_2Cu_3O_{7-x}$} superconducting thin
     films --- anisotropy and temperature dependence,''
\newblock {\em Phys. Rev. B}, vol. 50, no. 1, pp. 470, Jul 1994.



\bibitem{BAWillemsen95a}
Balam~A. Willemsen, John~S. Derov, Jos{\'e}~H. Silva, and S.~Sridhar,
\newblock ``Nonlinear response of suspended high temperature superconducting
  thin film microwave resonators,''
\newblock {\em IEEE Trans. Appl. Supercond.}, vol. 5, no. 2, pp. 1753--1755,
  June 1995.

\bibitem{CPBean62a}
C.~P. Bean,
\newblock ``Magnetization of hard superconductors,''
\newblock {\em Phys. Rev. Lett.}, vol. 8, pp. 250--253, 1962.

\bibitem{EZeldov94c}
E.~Zeldov, John~R. Clem, M.~M. McElfresh, and M.~Darwin,
\newblock ``Magnetization and transport currents in thin superconducing
  films,''
\newblock {\em Phys. Rev. B}, vol. 49, no. 14, pp. 9802--9822, April 1994.

\bibitem{EHBrandt93a}
Ernst~Helmut Brandt and Mikhail Indenbom,
\newblock ``Type-{II}-superconductor strip with current in a perpendicular
  magnetic field,''
\newblock {\em Phys. Rev. B}, vol. 48, pp. 12893--12906, 1993.

\bibitem{CampbellIEEE}
A.~M.~Campbell,
\newblock `` ''
\newblock {\em IEEE Trans. Appl. Supercond.}, vol. 5, pp. 687, 1995.


\end{thebibliography}

\end{document}